%% file: SSGluino_paper.tex
\documentclass[prl,superscriptaddress,twocolumn,showpacs,preprintnumbers,amsmath,amssymb]{revtex4}
\usepackage{subfigure}
\usepackage{epsfig}
\usepackage{dcolumn}
\usepackage{bm}
\usepackage{graphicx}
\usepackage{endnotes}
\usepackage{multirow}
\usepackage{xspace}
\usepackage{atlasphysics}
\usepackage[hyperindex,breaklinks]{hyperref}

\usepackage{preprintcover}  
\PreprintCoverPaperTitle{Search for gluinos in events with two same-sign leptons, jets and missing transverse momentum with the ATLAS detector \\ in $pp$ collisions at $\sqrt{s}=$7~TeV}  
\PreprintIdNumber{CERN-PH-EP-2012-061} 
\PreprintCoverAbstract{A search is presented for gluinos decaying via the supersymmetric partner of the top quark 
using events with two same-sign leptons, jets and missing transverse momentum. 
The analysis is performed with 
2.05~\ifb\ of integrated luminosity from $pp$ collisions at $\sqrt{s}=7$~TeV collected by the ATLAS detector at the LHC.
No excess beyond the Standard Model expectation is observed 
and exclusion limits are derived for simplified models where the gluino decays
via the supersymmetric partner of the top quark and in the MSUGRA/CMSSM framework. 
In those scenarios, gluino masses below 550~GeV are excluded at 95\% CL within the parameter space considered, 
significantly extending the coverage with respect to existing limits.
Depending on the model parameters, gluino masses up to 750~GeV can also be excluded at 95\% CL.}  
\PreprintJournalName{Physical Review Letters}

\newcommand{\tone}   {\tilde{t}^{}_{1}}
\newcommand{\tonebar}   {\tilde{t}^{*}_{1}}
\newcommand{\neut}  {\tilde{\chi}^{0}_{1}}  
\newcommand{\chipm}  {\tilde{\chi}^{\pm}_{1}} 
\def\etmiss{\ensuremath{E_{\mathrm{T}}^{\mathrm{miss}}}\xspace}

\def\mt{\ensuremath{m_{\mathrm{T}}}\xspace}

\newcommand{\gl}   {\tilde{g}}

\def\Wjets{$W$+jets\xspace}
\def\Zjets{$Z$+jets\xspace}

\renewcommand{\ttbar} {\ensuremath{t\bar{t}}\xspace}

\renewcommand{\met} {\ensuremath{E_{\mathrm{T}}^{\mathrm{miss}}}\xspace}
\renewcommand{\pt} {\ensuremath{p_\mathrm{T}}\xspace}

\def\met{\ensuremath{E_{\mathrm{T}}^{\mathrm{miss}}}}

\begin{document}
\title{Search for gluinos in events with two same-sign leptons, jets and missing transverse momentum with the ATLAS detector in $pp$ collisions at $\sqrt{s}=$7~TeV}
\author{The ATLAS Collaboration}
\affiliation{}
\begin{abstract} 

A search is presented for gluinos decaying via the supersymmetric partner of the top quark 
using events with two same-sign leptons, jets and missing transverse momentum. 
The analysis is performed with 
2.05~\ifb\ of integrated luminosity from $pp$ collisions at $\sqrt{s}=7$~\TeV\ collected by the ATLAS detector at the LHC.
No excess beyond the Standard Model expectation is observed 
and exclusion limits are derived for simplified models where the gluino decays
via the supersymmetric partner of the top quark and in the MSUGRA/CMSSM framework. 
In those scenarios, gluino masses below 550~GeV are excluded at 95\% CL within the parameter space considered, 
significantly extending the coverage with respect to existing limits.
Depending on the model parameters, gluino masses up to 750~GeV can also be excluded at 95\% CL.
\end{abstract}

\maketitle
\normalsize

Supersymmetry 
(SUSY)~\cite{Ramond:1971gb,Golfand:1971iw,Neveu:1971rx,Neveu:1971iv,Volkov:1973ix,Wess:1973kz,Wess:1974tw} 
is a theory beyond the Standard Model (SM) which predicts 
new bosonic partners for the existing fermions and fermionic partners for the
known bosons. 
In the framework of a generic $R$-parity conserving minimal 
supersymmetric extension of the SM (MSSM), SUSY 
particles are produced in pairs~\cite{Fayet:1977yc,Farrar:1978xj} and the lightest supersymmetric particle is stable, providing a possible candidate for dark matter.

In SUSY models, the gluino is a strongly-interacting Majorana fermion. Pair-produced 
gluinos therefore have equal probability to produce a pair of leptons that have the same charge (same-sign, SS) and the
opposite charge from their decays. 
The supersymmetric partner of the top quark (stop) has two mass eigenstates with $\tone$ being the lightest. 
Top quarks and $\tone$ squarks can be produced in the decay of the gluino
via $\gl\gl \to \ttbar \tone\tonebar, tt\tonebar\tonebar,\bar{t}\bar{t}\tone\tone$ 
~\cite{StopDecaysBeenakker:1997, HisanoKawagoeNojiri:2003, EWBaryoGenesis:2004,SDecay:2005, KramlRaklev:2006}.
The $\tone$ squark can further decay to the lightest chargino ($\chipm$) or lightest neutralino ($\neut$) 
via $\tone\to b\chipm$ or $\tone\to t\neut$ producing isolated leptons
in the semi-leptonic top quark decay or 
in the leptonic $\chipm$ decay and enriching the signal events 
with two or more leptons, jets and missing transverse momentum ($\met$) from the undetected neutralinos. 
This analysis considers events with a pair of isolated SS leptons, 
multiple high-$\pt$ jets and large $\met$. The requirement of SS leptons 
in the event suppresses the contribution from SM processes and thus
enhances the potential signal significance even for final state topologies with relatively soft jet kinematics. 
This article presents the first search for gluino-mediated stop production using the SS signature although other searches 
with SS leptons have been performed~\cite{2lepton35pb,2lepton1fb,ss_exotic_arxiv,PhysRevLett.98.221803}.
The analysis presented here complements the results of the ATLAS search based on single lepton plus $b$-jets~\cite{1lep_bjet_2fb}, enhancing the sensitivity in the experimentally difficult region near the kinematic limit for the production of two top quarks and a neutralino, and for the region with low stop masses.

ATLAS is a general-purpose detector~\cite{DetectorPaper:2008} at the LHC.
This search uses $pp$ collision data recorded from March to August 2011 
at a center-of-mass energy of 7 \TeV. 
The data 
set corresponds to a total integrated luminosity of $2.05 \pm 0.08$~fb$^{-1}$~\cite{Aad:2011dr,lumiCONF} after the application of 
data quality requirements. 
Events are selected using single lepton and dilepton triggers that have constant efficiency as a function of lepton $\pt$ 
above the offline $\pt$ cuts used in the analysis.

Jets are reconstructed from three-dimensional calorimeter 
energy clusters by using the anti-$k_t$ jet algorithm~\cite{Cacciari:2008gp, Cacciari200657} 
with a radius parameter of 0.4. 
The measured jet energy is corrected 
for inhomogeneities in, and for the non-compensating nature of, the calorimeter 
by using $\pt$- and $\eta$-dependent correction factors~\cite{JES}. 
Only jet candidates with $\pt > 20$~GeV 
within $|\eta| < 2.8$ are retained. Events with any jet that
fails the jet quality criteria designed to remove noise and
non-collision backgrounds~\cite{JES} are rejected.

Electron candidates are required to have $\pt > 20$~\GeV, $|\eta| < 2.47$ 
and satisfy the `tight' selection criteria defined in Ref.~\cite{ELE_arxiv}. They are also required to be isolated: the  scalar sum of \pt of tracks within a cone in the $\eta-\phi$ plane of radius 
$\Delta R = 0.2$ around the candidate excluding its own track, $\Sigma_{\pt}$, must be less than 10\% of the electron \pt.
Muon 
candidates are required to have $\pt > 20$~GeV, $|\eta| < 2.4$ and are 
identified by matching an extrapolated inner detector track and one or more 
track segments in the muon spectrometer~\cite{MuonRecoEff}. They must have longitudinal and transverse impact parameter within 1~mm and 0.2~mm 
of the primary vertex~\cite{PrimVertex}, respectively, and $\Sigma_{\pt}< 1.8$~GeV. 

The calculation of $\etmiss$~\cite{MET_arxiv} is based on the vectorial sum 
of the $\pt$ of the reconstructed jets (with $\pt > 20 \GeV$ and $|\eta| < 4.5$), 
leptons and the calorimeter energy clusters 
not belonging to reconstructed objects.

During part of the data-taking period, a localized electronics failure in the 
electromagnetic calorimeter created a dead region ($\Delta \eta \times \Delta \phi \simeq 1.4 \times 0.2 $). 
For jets in this region, a correction to their energy is made using the energy
depositions in the neighbouring cells, and is propagated to $\met$.  If this correction 
projected onto the direction of the $\met$ is larger than 10~GeV or 10\% of the $\met$, 
the event is discarded~\cite{1leptonPRD}. Events with reconstructed electrons in the calorimeter dead region are also rejected.

Events in which the two highest-$\pt$ leptons ($\ell=e$, $\mu$) have the 
same charge and with at least four jets with $\pt > 50$ \GeV\ are selected. In addition, two signal regions are considered:
SR1, which requires $\met > 150$~\GeV; and SR2, which in addition requires $\mt > 100$~\GeV,
where $\mt$  is the transverse mass of the $\met$ and the highest-$\pt$ 
lepton defined as $\mt^2=2 p_T^{\ell}\met(1 - \cos(\Delta\phi(\ell,\met))$. 
This final $\mt$ cut helps reducing the $\ttbar$ background. 
The signal regions are optimized based on several models where SS dileptons are produced in gluino decays. 
In signals such as the MSUGRA/CMSSM (minimal supergravity or constrained minimal supersymmetric standard model)~\cite{msugra,Kane:1993td}, the directions of the lepton and $\tilde{\chi}_1^0$ are strongly correlated as they originate from the decay of a common parent particle (usually $\tilde{\chi}_1^\pm$ or the next-to-lightest neutralino, $\tilde{\chi}_2^0$). This leads to a softer $\mt$ spectrum than that found in gluino-mediated stop signal models, where the lepton and the $\tilde{\chi}_1^0$ originate from different parent particles and are thus uncorrelated.

Simulated Monte Carlo (MC) event samples are used to aid in the description of the background and 
to model the SUSY signal. 
Top-quark pair and single-top production are simulated with {\tt MC@NLO}~\cite{mcatnlo}, 
fixing the top-quark mass at 172.5~GeV, and using the next-to-leading-order (NLO) parton density function (PDF) set 
{\tt CTEQ6.6}~\cite{cteq66}. Samples of \Wjets\ and \Zjets with both light- 
and heavy-flavor jets 
are generated with 
{\tt ALPGEN}~\cite{alpgen} and PDF set {\tt CTEQ6L1}~\cite{cteq6l1}.  
The fragmentation and hadronization for the {\tt ALPGEN} and {\tt MC@NLO} samples are 
performed with {\tt HERWIG}~\cite{herwig}, using {\tt JIMMY}~\cite{jimmy} for the 
underlying event.
Samples of $\ttbar Z$, $\ttbar W$ and $\ttbar WW$ (referred to as $\ttbar + X$) are generated with {\tt MADGRAPH}~\cite{madgraph} 
interfaced to {\tt PYTHIA}~\cite{pythia}. The total LO cross section for these samples is 0.39 pb and is normalized to NLO using a $K$-factor 
of 1.3~\cite{PetrielloPriv}.
Diboson samples are generated with {\tt HERWIG} for $W^{\pm}W^{\mp}$, $WZ$ and $ZZ$ 
processes and with {\tt MADGRAPH} for $W^{\pm}W^{\pm} qq$ processes. 
The total NLO cross section for the diboson background is 71 pb~\cite{MCFM_PRD, MCFM_JHEP}.  
SUSY signal processes are simulated for various models using {\tt HERWIG++}~\cite{herwig} v2.4.2. 
The SUSY sample yields are normalized to the results of NLO calculations, as obtained using {\tt PROSPINO}~\cite{Beenakker:1997ut} 
v2.1. The {\tt CTEQ6.6M}~\cite{cteq6m} parameterization of the PDFs is used. 
The tunings of the MC parameters of Ref.~\cite{atlastune} are used in the production of the MC samples, which are 
processed through a detector simulation~\cite{atlassimulation} based on {\tt GEANT4}~\cite{geant4}. 
Effects of multiple proton-proton interactions per bunch crossing are included in the simulation.

The SM backgrounds are evaluated using a combination of MC simulation and data-driven techniques. 
SM processes that generate events containing jets which are misidentified as leptons or where a lepton from a $b$- or $c$-hadron decay is selected are 
collectively referred to as ``fake-lepton'' background. It generally consists of semi-leptonic  $\ttbar$, single top, $W$+jets and strong light- and heavy-flavor jet production. 
The contribution from the ``fake-lepton'' background is estimated from data with a method similar to that described in Ref.~\cite{top1,top2} by loosening the lepton identification and isolation criteria.  
For electrons the `medium' criteria are used instead of the `tight' criteria~\cite{ELE_arxiv}, and for both electrons and muons the isolation criterion is relaxed. 
The method counts the number of observed events containing loose-loose, loose-tight, tight-loose and tight-tight lepton pairs. 
The probability of loose real leptons passing the tight selection criteria is obtained using a $Z\to\ell^+\ell^-$ sample. The probability of loose fake leptons to pass the tight selection criteria is determined as a function of the lepton $\pt$ using multijet control samples obtained by requiring two SS leptons and low $\met$. Using these probabilities, relations are obtained for the observed event counts in the signal regions as functions of the numbers of events containing fake-fake, fake-real, real-fake and real-real lepton pairs. 
These can be solved simultaneously to estimate the number of background events~\cite{top1,top2}. 
The results of the estimations have been validated with data in control regions obtained by reversing the $\met$ or jet multiplicity cuts used in the signal regions.

Background events from charge misidentification (dominated by electrons which have undergone 
hard bremsstrah\-lung with subsequent photon conversion) are estimated using a partially data-driven 
technique~\cite{2lepton1fb}. 
The probability of charge misidentification is calculated from MC and corrected by consideration of the number of events in data with SS electron pairs 
and invariant mass within 15~GeV of the $Z$-boson mass. 
This probability is applied to $\ttbar$ MC events producing $e^\pm\ell^\mp$ pairs to evaluate the number of SS events from incorrect charge assignment in each signal region. 
The probability of misidentifying the charge of a muon and the contributions in the signal regions from charge misidentification of $Z/\gamma^*$+jets and other SM backgrounds are negligible.

Contributions from other SM background sources (diboson and $\ttbar + X$) 
are evaluated using the MC samples described above. In these processes, real SS lepton pairs are produced and their contribution to the signal regions can be described with MC. 
In particular, the contribution from the experimentally unmeasured $\ttbar + X$ processes has been studied using several MC generators.
The background from cosmic rays is evaluated with data using the method in Ref.~\cite{2lepton1fb} and its contribution is negligible in the signal regions.

Systematic uncertainties are estimated in the signal regions for the background
and the SUSY signal processes. The primary sources of
systematic uncertainties in the background are the jet energy scale calibration (35\%), the 
jet energy resolution (10\%), uncertainties on lepton and jet reconstruction and identification (5\%), MC modeling and theoretical cross section uncertainties (40\%-70\%). 
In particular, the theoretical uncertainties on the cross section of the $\ttbar +X$ processes are found 
to be between 35\%-55\% by varying factorization and renormalization 
scales and 25\% due to PDF uncertainties. In addition, a 50\% uncertainty is assigned
on the $K$-factor used to obtain the NLO cross section~\cite{PetrielloPriv}.
In the fake-lepton background estimation systematic uncertainties are assigned to 
the probabilities for loose fake leptons to pass the tight selection. This accounts 
for potentially different compositions of the signal and control regions. These uncertainties 
vary in the 10\%-80\% range depending on the lepton $\pt$ and are evaluated using data samples with jets of different energies.
The absolute uncertainty for each background source is given in Table~\ref{tab:FinalResultsSR}. 
Systematic uncertainties on the signal expectations are evaluated 
through variations of the factorization and renormalization scales 
between half and twice their default values, and by 
including the uncertainty on $\alpha_s$ and on the PDF provided by {\tt CTEQ6}. 
Uncertainties are calculated for individual SUSY processes. The total uncertainty
varies in the 20\%-40\% range for the considered MC signals. Any correlations of the systematic
uncertainties in signals and background are taken into account.

\vspace*{-0.2cm}
\begin{table}[htp!]
    \renewcommand\arraystretch{1.1}
  \caption{Number of expected SM background events together with the number of observed events in data. 
    The errors are a combination of the uncertainties due to MC statistics, statistical uncertainties in control regions and systematic uncertainties. 
    Observed and expected upper limits at 95\% confidence level on $\sigma_{\rm vis}=\sigma\times\epsilon\times A$, together with the
    $\pm 1\sigma$ errors on the expected limits are also shown.} 
  \label{tab:FinalResultsSR}
   \centering
 \begin{tabular}{l|cc}
   \hline\hline
     & SR1 & SR2 \\
   \hline
   \ttbar+X  & 0.37 $\pm$ 0.26 & 0.21 $\pm$ 0.16 \\
   Diboson  & 0.05 $\pm$ 0.02 & 0.02 $\pm$ 0.01 \\
   Fake-lepton  & 0.34 $\pm$ 0.20 & $<0.17$ \\
      Charge mis-ID  & 0.08 $\pm$ 0.01 & 0.039 $\pm$ 0.007 \\
   Total SM &  0.84 $\pm$ 0.33 & 0.27 $\pm$ 0.24 \\\hline
   Observed & 0 & 0 \\\hline\hline
   $\sigma_{\rm vis}^{\rm obs}$ [fb] & $<$1.6  & $<$1.5  \\ 
   $\sigma_{\rm vis}^{\rm exp}$ [fb] & $<$1.7$^{+0.5}_{-0.1}$ & $<$1.6$^{+0.2}_{-0.1}$ \\
  \hline\hline
  \end{tabular}

\end{table}

\begin{figure}[h!]
\centering                                                            
  \includegraphics[scale=0.35]{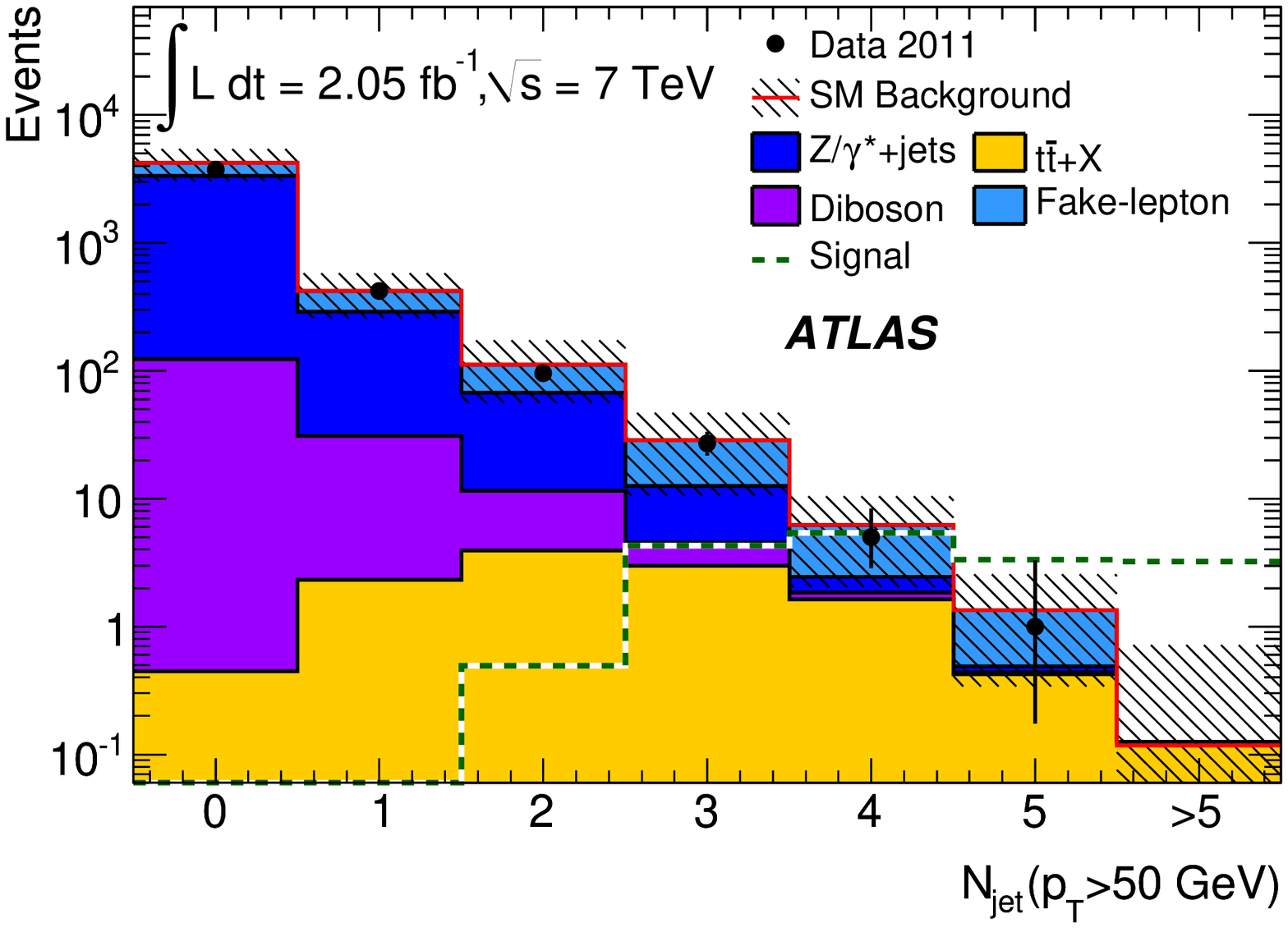}\\
  \includegraphics[scale=0.35]{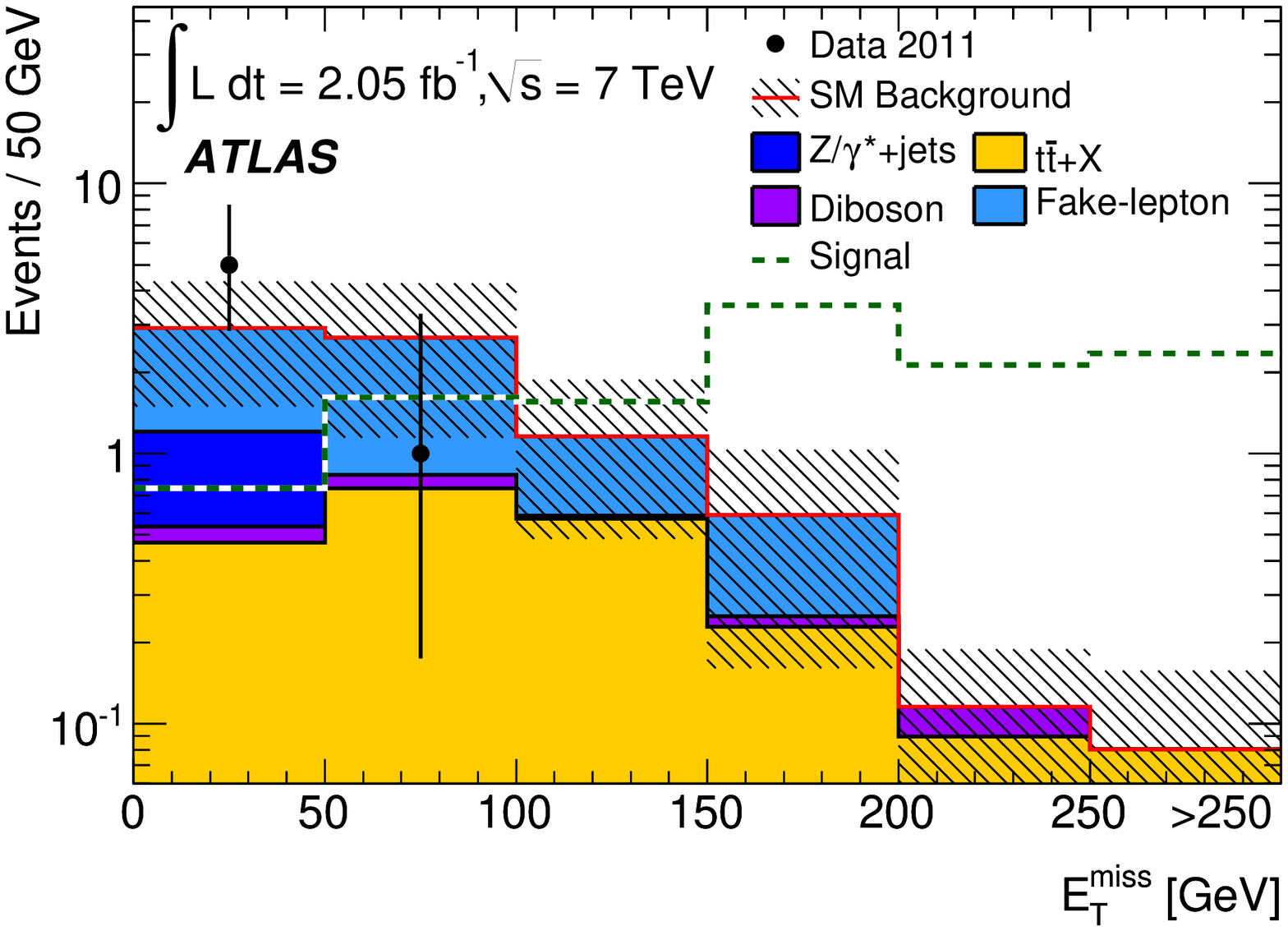}
\vspace{-.2cm}
 \caption{Number of jets with $\pt>50$~GeV for events with 2 SS leptons (top) and the $\met$ distribution for events with 2 SS leptons and at least 4 jets with $\pt>50$~GeV (bottom). Errors on data points are statistical, while the error band on the SM background represents the total uncertainty. The component labelled ``Fake-lepton'' is evaluated using data as described in the text. The component labelled ``$Z$/$\gamma^*$+jets'' is estimated from MC. No  estimation of the charge mis-ID is included in the distribution.  
 The component labelled ``Signal'' corresponds to a signal obtained with the decay $\tilde{g}\to\ttbar\tilde{\chi}_1^0$ via off mass-shell $\tilde{t}$ ($m_{\tilde{t}} = 1.2$ TeV) in $\tilde{g}\tilde{g}$ pair-produced events with  $m_{\tilde{g}}=650$~GeV and $m_{\tilde{\chi}_1^0}=150$~\GeV. }
 \label{fig:res_plots}
\vspace{-0.5cm}
\end{figure}

Figure~\ref{fig:res_plots} shows the distribution of the number of jets with $\pt>50$~\GeV\ for events with 2 SS leptons, and the $\met$ distribution for events with 2 SS leptons and at least four jets with $\pt>50$~\GeV. The contributions from all the SM backgrounds are shown together with their total statistical and systematic uncertainties. For illustration, the distribution for a signal obtained with the decay $\tilde{g}\to\ttbar\tilde{\chi}_1^0$ in $\tilde{g}\tilde{g}$ pair-produced events with $m_{\tilde{g}}=650$~GeV and $m_{\tilde{\chi}_1^0}=150$~GeV is also shown. The data are in agreement with the SM background expectation and once four jets of $\pt>50$~\GeV\ are required no event is observed with $\met>150$~\GeV.

Table~\ref{tab:FinalResultsSR} shows the number of expected events 
in the signal regions for each background source together with the observed number of events. 
The expectation from the SM is estimated to be 
less than one event for each signal region with no events observed in data. Limits at 95\% confidence level (CL) are derived on the visible cross section $\sigma_{\rm vis} = \sigma\times\epsilon\times A$ where $\sigma$ is the total production 
cross section for any new signal producing SS dileptons, $A$ is the acceptance defined by the fraction of events passing geometric 
and kinematic cuts at particle level and $\epsilon$ is the detector reconstruction, identification and trigger efficiency. For the signal shown in Figure~\ref{fig:res_plots}, the acceptance and efficiency are $1.5\%$ and $55\%$, respectively.
Limits are set using the $CL_s$ prescription, as described in Ref.~\cite{Read:2002hq}. The results are given in Table~\ref{tab:FinalResultsSR}.

\begin{figure}[htb!]
  \begin{center}
    \includegraphics[scale=0.4]{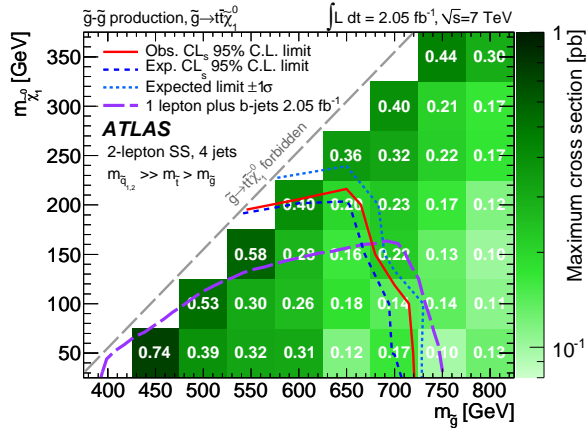}
    \end{center}
\vspace{-.5cm}
  \caption{Expected and observed 95\% CL 
exclusion limits in the $\tilde{g}\to\ttbar\tilde{\chi}_1^0$ (via off mass-shell $\tilde{t}$, $m_{\tilde{t}} = 1.2$ TeV) simplified model as a function 
of the gluino and neutralino masses, together with existing limits~\cite{1lep_bjet_2fb}. The $-1\sigma$ limit lies outside the range of the figure. The production cross section upper limits at 95\% CL are also shown.}
\label{fig:excl_sbLSP_dphi32}
\end{figure}

The results obtained in SR2 are interpreted in a simplified model where gluinos are only produced in pairs, the stop ($m_{\tilde{t}}=1.2$ TeV)
is heavier than the gluino, and only the gluino three-body decay $\tilde{g}\to\ttbar\tilde{\chi}_1^0$ via an off-shell stop is allowed. 
Figure~\ref{fig:excl_sbLSP_dphi32} shows the limit 
in the gluino-neutralino mass plane. 
For a gluino mass of 650 GeV, neutralino masses below 215 GeV are excluded at 95\% CL. 
For a neutralino mass of 100 GeV, gluino masses below 715 GeV are similarly excluded.
The $-1 \sigma$ uncertainty limit on the expected limit lies outside the range of the 
figure as a consequence of the low number of expected signal events and a total signal uncertainty that reaches close to 50\%. The results can be generalised in terms of production cross section upper limits at 95\% CL for $\tilde{g}\tilde{g}$ pair production processes with the produced particles decaying into $\ttbar\tilde{\chi}_1^0$ final states, as also shown in Figure~\ref{fig:excl_sbLSP_dphi32}. 

\begin{figure}[h!]
  \begin{center}
    \includegraphics[scale=0.4]{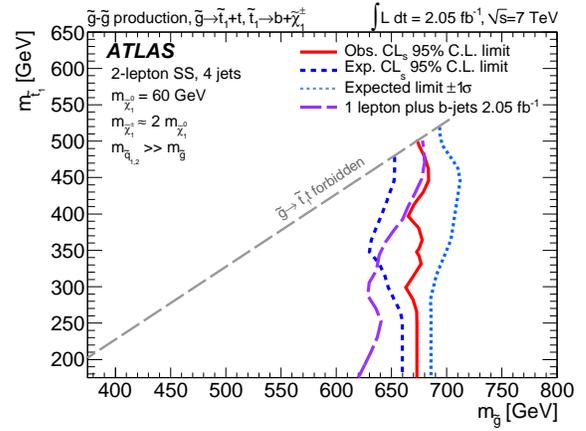}
        \end{center}
\vspace{-.5cm}
  \caption{Expected and observed 95\% CL exclusion limits in the $\tilde{g}\to\tilde{t}_1t$ with $\tilde{t}_{1}\to b\tilde{\chi}_{1}^{\pm}$ model 
  as a function of the gluino and stop masses assuming that $m_{\tilde{\chi}^\pm_1}\approx 2m_{\tilde{\chi}^0_1}$. The $-1\sigma$ band lies outside the range of the figure.}
\label{fig:excl_Gtst}
\end{figure}

The results in SR2 are also interpreted considering gluino pair production 
followed by the $\tilde{g}\to\tilde{t}_1t$ decay. Only stop decays 
via $\tilde{t}_{1}\to b\tilde{\chi}_{1}^{\pm}$ are considered with 
$m_{\tilde{\chi}^\pm_1} \approx 2m_{\tilde{\chi}^0_1}$ and $m_{\tilde{\chi}^0_1}=60$~\GeV. 
Figure~\ref{fig:excl_Gtst} shows the exclusion limit as a function 
of gluino and stop masses, where gluino masses below 660~GeV are excluded  at 95\% CL
for stop masses below 460~\GeV.

\begin{figure}[h!]
  \begin{center}
    \includegraphics[scale=0.4]{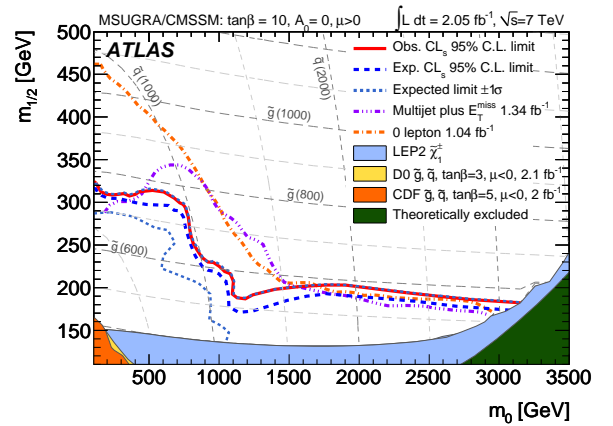}
    \end{center}
\vspace{-.5cm}
  \caption{Expected and observed 95\% CL exclusion limits in the MSUGRA/CMSSM $(m_{0},m_{1/2})$ plane for $\tan\beta=10$, $A_0=0$ and $\mu>0$, compared to existing limits~\cite{Aaltonen:2008rv,:2007ww,:2010d0sb,lep2,MultijetATLAS,0leptonATLAS}. 
}
\label{fig:excl_mSUGRA}
\end{figure}

The results in SR1 are interpreted within the MSUGRA-CMSSM framework 
in terms of limits on the universal scalar and gaugino mass parameters $m_0$ and $m_{1/2}$,
 as shown in Figure~\ref{fig:excl_mSUGRA}. These are presented for fixed values of the universal 
trilinear coupling parameter $A_0=0$, ratio of the vacuum expectation values of the two 
Higgs doublets $\tan\beta=10$, and Higgs mixing parameter $\mu > 0$. In this model, values of 
$m_{1/2}$ below 300 GeV are excluded at 95\% CL for $m_0$ values below 750 GeV, 
and $m_{1/2}$ values below 180 GeV are excluded over the entire $m_0$ region considered.
These are equivalent to the exclusion of gluino masses below $\sim$550~GeV independent of the 
squark mass (and gluino masses below $\sim$750~GeV for squark masses below 1~\TeV).

In summary, a search for SUSY with two SS leptons, jets and missing 
transverse momentum has been performed using 2.05~fb$^{-1}$ of ATLAS data.
With no events observed in the signal regions, limits have been derived 
in the context of models where top quarks are produced in gluino decays and MSUGRA/CMSSM scenarios. 
In all these signal models, gluino masses below $\sim$550~GeV are excluded at 95\% CL
within the parameter space considered
and gluino masses up to $\sim$750~GeV are excluded at 95\% CL, depending on the model parameters. 
The results of this analysis are complementary to and extend the current exclusion limits on the gluino mass beyond those from other ATLAS searches~\cite{1lep_bjet_2fb,MultijetATLAS,0leptonATLAS}.

We thank CERN for the very successful operation of the LHC, as well as the
support staff from our institutions without whom ATLAS could not be
operated efficiently.

We acknowledge the support of ANPCyT, Argentina; YerPhI, Armenia; ARC,
Australia; BMWF, Austria; ANAS, Azerbaijan; SSTC, Belarus; CNPq and FAPESP,
Brazil; NSERC, NRC and CFI, Canada; CERN; CONICYT, Chile; CAS, MOST and NSFC,
China; COLCIENCIAS, Colombia; MSMT CR, MPO CR and VSC CR, Czech Republic;
DNRF, DNSRC and Lundbeck Foundation, Denmark; ARTEMIS and ERC, European Union;
IN2P3-CNRS, CEA-DSM/IRFU, France; GNAS, Georgia; BMBF, DFG, HGF, MPG and AvH
Foundation, Germany; GSRT, Greece; ISF, MINERVA, GIF, DIP and Benoziyo Center,
Israel; INFN, Italy; MEXT and JSPS, Japan; CNRST, Morocco; FOM and NWO,
Netherlands; RCN, Norway; MNiSW, Poland; GRICES and FCT, Portugal; MERYS
(MECTS), Romania; MES of Russia and ROSATOM, Russian Federation; JINR; MSTD,
Serbia; MSSR, Slovakia; ARRS and MVZT, Slovenia; DST/NRF, South Africa;
MICINN, Spain; SRC and Wallenberg Foundation, Sweden; SER, SNSF and Cantons of
Bern and Geneva, Switzerland; NSC, Taiwan; TAEK, Turkey; STFC, the Royal
Society and Leverhulme Trust, United Kingdom; DOE and NSF, United States of
America.

The crucial computing support from all WLCG partners is acknowledged
gratefully, in particular from CERN and the ATLAS Tier-1 facilities at
TRIUMF (Canada), NDGF (Denmark, Norway, Sweden), CC-IN2P3 (France),
KIT/GridKA (Germany), INFN-CNAF (Italy), NL-T1 (Netherlands), PIC (Spain),
ASGC (Taiwan), RAL (UK) and BNL (USA) and in the Tier-2 facilities
worldwide.

\bibliography{SSGluino_Paper}
\onecolumngrid
\clearpage
\input{atlas_authlist}

\end{document}

%% file: atlas_authlist.tex
\begin{flushleft}
{\Large The ATLAS Collaboration}

\bigskip

G.~Aad$^{\rm 48}$,
B.~Abbott$^{\rm 110}$,
J.~Abdallah$^{\rm 11}$,
S.~Abdel~Khalek$^{\rm 114}$,
A.A.~Abdelalim$^{\rm 49}$,
A.~Abdesselam$^{\rm 117}$,
O.~Abdinov$^{\rm 10}$,
B.~Abi$^{\rm 111}$,
M.~Abolins$^{\rm 87}$,
O.S.~AbouZeid$^{\rm 157}$
% ...
\bigskip

% ...
$^{10}$ Institute of Physics, Azerbaijan Academy of Sciences, Baku, Azerbaijan\\
$^{11}$ Institut de F\'isica d'Altes Energies and Departament de F\'isica de la Universitat Aut\`onoma  de Barcelona and ICREA, Barcelona, Spain\\
% ...
$^{48}$ Fakult\"{a}t f\"{u}r Mathematik und Physik, Albert-Ludwigs-Universit\"{a}t, Freiburg i.Br., Germany\\
$^{49}$ Section de Physique, Universit\'e de Gen\`eve, Geneva, Switzerland\\
% ...
$^{87}$ Department of Physics and Astronomy, Michigan State University, East Lansing MI, United States of America\\
% ...
$^{110}$ Homer L. Dodge Department of Physics and Astronomy, University of Oklahoma, Norman OK, United States of America\\
$^{111}$ Department of Physics, Oklahoma State University, Stillwater OK, United States of America\\
% ...
$^{114}$ LAL, Univ. Paris-Sud and CNRS/IN2P3, Orsay, France\\
% ...
$^{117}$ Department of Physics, Oxford University, Oxford, United Kingdom\\
% ...
$^{157}$ Department of Physics, University of Toronto, Toronto ON, Canada\\
% ...
\end{flushleft}
